\documentclass[a4paper,fleqn,usenatbib]{mnras}

\usepackage{newtxtext,newtxmath}

\usepackage[T1]{fontenc}
\usepackage{ae,aecompl}

\usepackage{graphicx}	
\usepackage{amsmath}	
\usepackage{amssymb}	
\usepackage{flafter}
\usepackage{float}
\usepackage{threeparttable}

\title[]{The ionic composition of the local absorber towards 3C 273}

\author[Gissis et al.]{
Itay Gissis,$^{1}$
Uria Peretz,$^{1}$
Ehud Behar$^{1}$
\\
$^{1}$Technion, Inst. of Technology, Israel 320003\\
}

\date{Accepted 2020 January 17. Received  2020 January 17; in original form 2019 November 15.}

\pubyear{2019}

\begin{document}
\label{firstpage}
\pagerange{\pageref{firstpage}--\pageref{lastpage}}
\maketitle

\begin{abstract}
Hot ionized gas is observed in the local vicinity of our galaxy through spectral absorption features. The most common hypothesis is that this gas forms a halo surrounding our Milky-Way (MW), in collisional ionization equilibrium. In this paper we investigate the elemental abundance of this hot and ionized local gas. We use a 2.4 Ms stacked X-ray spectrum of the bright blazar 3C 273 and probe the local absorption features. Using ion-by-ion fitting of the X-ray absorption lines we derive the column density of each ionization species. Based on the column densities we reconstruct the Absorption measure distribution (AMD), namely the hydrogenic column density as a function of temperature. We report the elemental abundances of C, N, Ne, and Fe relative to solar O. Previous measurements of local X-ray emission lines in conjunction with the present column densities indicate a scale height of $1-80$ kpc and hydrogen number density of $10^{-4}-10^{-3}$cm\textsuperscript{-3} for the hot ionized gas. Additionally, we detect He-like O lines from the quasar broad line region with velocities of 6400$\pm$1500 km\,s$^{-1}$.
\end{abstract}

\begin{keywords}
ISM: abundances -- quasars: supermassive black holes -- X-rays: general
\end{keywords}



\section{Introduction}

From the very first AGN spectra obtained with the Chandra and XMM-Newton observatories, absorption features at z=0 were detected. The most significant absorption features are the resonances of O\textsuperscript{+6} and to a lesser extent O\textsuperscript{+7}, which imply hot temperature of $\sim10^6$ K and above. See review by \cite{Paerels&Kahn2003}. These lines are detected primarily using AGNs as back-illuminators \citep{Nicastro2002,Fang2002, Rasmussen2003, Wang2005, Williams2005, Fang2006, Yao&Wang2007, Hagihara2010}, but also via the soft X-ray background (SXRB) \citep{Snowden1997, McCammon2002, Yoshino2009, Henley&Shelton2010, Henley&Shelton2012}. \\
The origin of this local hot absorber is debated. Researchers ascribe it to an extended galactic halo \citep{Wang2005, Fang2006} or the intergroup medium of the Local Group \citep{Nicastro2002, Williams2005}. \cite{Bregman&Lloyd-Davies2007} studied O\textsuperscript{+6} absorption in 25 lines of sight and concluded that these lines likely originate in the MW halo. \cite{Fang2015} expanded the survey to 43 lines of sight also concluding that the O\textsuperscript{+6} absorption  originates from local MW hot gas. Their strongest support for this conclusion was the uniformity and isotropy of the lines Equivalent Width (EW) in all galactic directions which implies an absorber presumably in the form of a spherical halo. Hot gaseous halos are known to exist in other galaxies as well \citep{Bregman&Houck1997, Li2008, Anderson&Bregman2010, Anderson&Bregman2011}, which strengthens even more the plausibility of a MW halo origin.
Seeking the size and profile of this hot halo, large surveys in different galactic directions were carried out. \cite{Miller&Bregman2015} examined emission measures from $\sim$650 sightlines. They modeled the density profile of the halo as a power-law with $n(r)\sim r^{-1.5}$, also estimating its total mass up to 250 kpc to be $4.3^{+0.9}_{-0.8}\times10^{10}M_\odot$, assuming a metallicity of $Z\geqslant0.3\,Z_\odot$. This accounts for $\lesssim50\%$ of the MW missing baryons \citep{Miller&Bregman2015}.

A measurement of the thermal and chemical structure of the halo is still lacking. A notable exception are the three spectroscopic studies of Mrk 421 \citep{Williams2005}, Mrk 271 \citep{Williams2006} and PKS 2155-304 \cite{Williams2007} that tried to reconcile the O\textsuperscript{+6} measurement with those of O\textsuperscript{+5} in the UV, while trying to determine whether the absorbing plasma is hot and collisionally ionized in the MW halo, or colder photoionized inter galactic gas. \cite{Nevalainen2017} identified O\textsuperscript{+4} and O\textsuperscript{+3} towards PKS 2155-304, which is evidence for a cooler temperature gas at $\log(T[\mathrm{K}])=5.2$. This led them to favor collisional ionization for this gas. For Mrk 421 \cite{Williams2005} also published elemental abundances of C/O, N/O, Ne/O, which were weakly constrained ($\sim0.25-4)$. 

The present work focuses on the blazar 3C 273 located at redshift z=0.158 \citep{Schmidt1963}, galactic longitude $l=289.95\degr$ and latitude $b=64.36\degr$. 3C 273 is interesting as the EW of its z=0 O\textsuperscript{+6} absorption line is high, and its broadening is beyond thermal \citep{Fang&Xiaochuan2014}. Earlier on, its X-ray spectrum was studied by \cite{Fang2002}, who investigated the ionizing source of the absorbing gas and found it to be collisionally ionized and in the galaxy, but could not rule out a photoionized inter-group absorber. \cite{Rasmussen2003} studied the oxygen absorption along the sightlines towards 3C 273, Mrk 421 and PKS 2155-304, combined with emission measure from \cite{McCammon2002}. They derived an upper limit on the electron density of $n_e<2\times10^{-4}\mathrm{cm}^{-3}$ and a lower limit on the scale length of $L<140$ kpc (for $\log(T[\mathrm{K}])=6.3-6.4$), placing the absorber in the local intergroup medium. More recently, \cite{Fang&Xiaochuan2014}, who use XMM-Newton (769 ks total exposure time) and Chandra (468 ks) spectra of 3C 273, report the EW of O\textsuperscript{+7} and Ne\textsuperscript{+8} as well. They used these ions and the UV measured O\textsuperscript{+5}, assuming they arise from the same gas, to find the temperature of the collisionally ionized absorber to be 1.5-1.8$\times$10\textsuperscript{6} K. Using the measurements of the Soft X-ray Background by \cite{Henley&Shelton2012} they estimate the distance of the O\textsuperscript{+6} gas to be 5-15 kpc, and ascribe it to the far Galactic disk.

\section{Analysis} \label{sec:Analysis}
\subsection{Data Reduction} \label{subsec:DataReduction} 
In the present work, we analyze XMM-Newton RGS observations to study the local absorber towards 3C 273. We use the XMM-Newton science analysis system \citep[SAS,][]{Gabriel2004}\footnote{https://www.cosmos.esa.int/web/xmm-newton/sas-installation} to co-add RGS1 and RGS2 spectra for a total of 2.4 Ms exposure time and $3.8\times10^6$ source photons. Standard pipeline data reduction process, including background subtraction was applied to each observation prior to stacking. We analyze the 8 \AA{} -- 36 \AA{} band where the RGS is most sensitive.

\begin{figure*}
	\includegraphics[width=\textwidth,height=10cm]{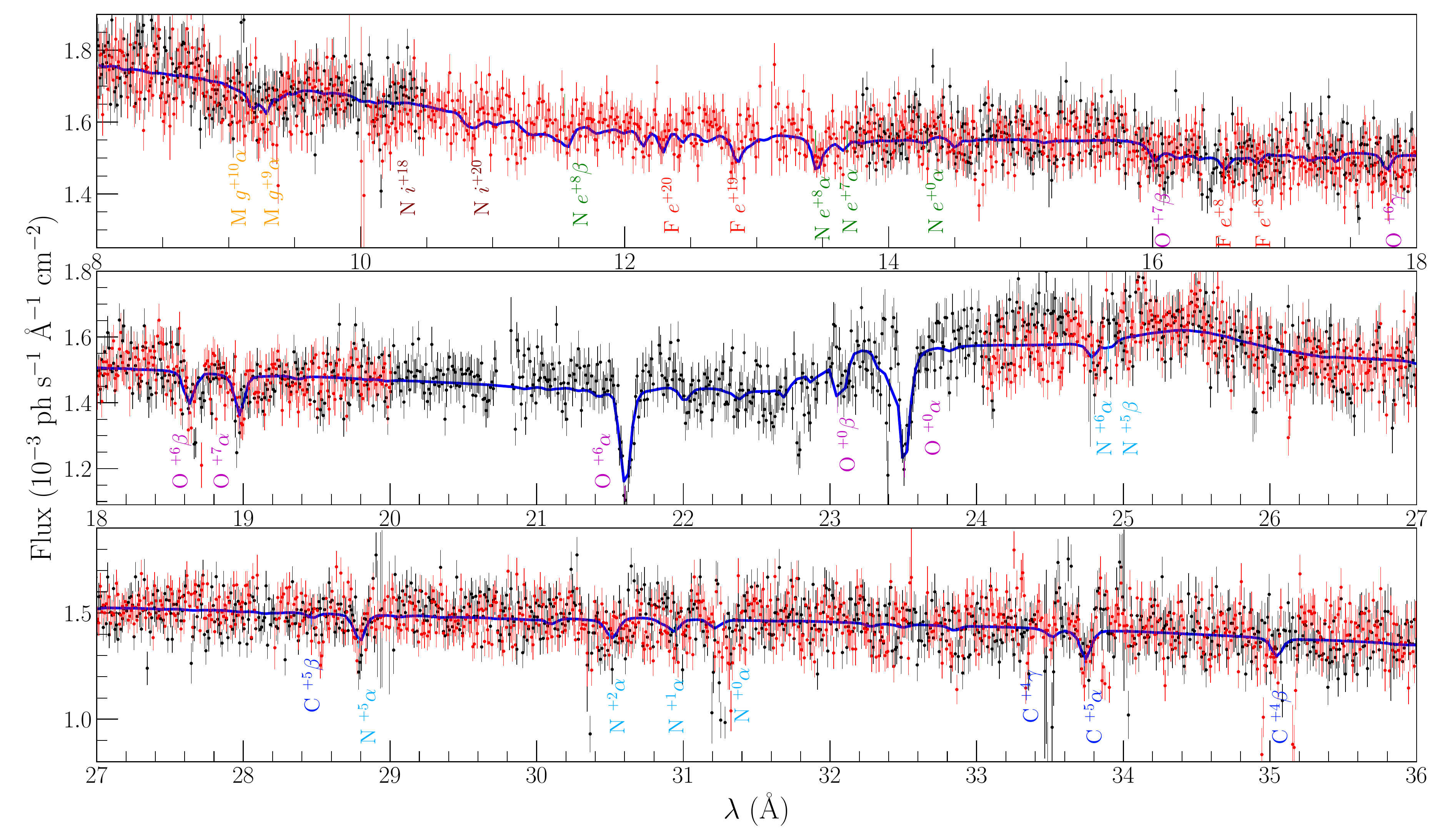}
	\caption{Stacked RGS1 (black) and RGS2 (red) spectra of 3C 273. The solid blue line is the best fit model. The strongest lines are labeled on the plot.}
	\label{fig:spectrum}
\end{figure*}

In Figure  \ref{fig:spectrum} we plot the stacked spectra of RGS1 and RGS2 along with our best fit model (see Section \ref{subsec:SpectralFitting} below). Line features, such as the O\textsuperscript{+7} at 18.97\AA{}, O\textsuperscript{+6} at 18.63\AA{}, N\textsuperscript{+6} at 24.78\AA{}, N\textsuperscript{+5} at 28.79\AA{}, C\textsuperscript{+5} at 33.73\AA{} and even the Mg K$\alpha$ lines at 9.17\AA{} and 9.30\AA{} are prominent in both spectra. The broad emission feature around 24-25\AA{} is also apparent. Notice that instrumental features mostly due to the RGS chip gaps, e.g. at 10\AA{}, 13.05\AA{}, 21.8\AA{}, 23.35\AA{}, 29\AA{}, and 33.5\AA{} are not spectral lines.

Our goal of temperature and abundance measurements, relies on the global absorption properties of the spectrum, weighted mostly towards the strong absorption lines, where the uncertainties are small. Since they are attributed to the local absorber, their S/N is improved by co-adding observations regardless of possible fluctuations is the source spectrum. We do not include Chandra observations in our analysis, as its S/N is inferior to that of the RGS. However. we did fit the oxygen and neon line EWs in the LETG 1\textsuperscript{st} order combined spectrum and found it to agree with the RGS results.

\subsection{Ion-By-Ion Fitting} \label{subsec:SpectralFitting}
The absorption spectrum of 3C 273 indicates both neutral and ionized gas components along the line of sight. For the colder, i.e. atomic and mildly ionized species (up to +3) we use the ISMAbs model \citep{Gatuzz2014}. The ionized component, is modeled using our ion-by-ion fitting model - Ibifit \citep{Peretz&Behar2017}. The model fits the column density of all the astrophysical abundant ions, taking into account the transitions, oscillator strengths and photoelectric edges, which were computed using the HULLAC code \citep{Barshalom2001}. The width of the absorption lines is unresolved by the RGS ($\Delta\lambda_\mathrm{RGS}\approx0.07$\AA{}), but the fit constrains it to a turbulent velocity (Doppler $b$ parameter) of 100$\pm50$ km\,s$^{-1}$, which is of the order of the UV line width (see Section\,\ref{sec:UVLineOfSight}).

The spectrum of 3C 273 comprises a continuum modeled as a power-law and a black-body as well as broad lines modeled using a photo-ionized emission component. Global fitting has the advantage of constraining all lines and edges of each ion together, but its continuum may deviate locally from the data.
The most dominant broad feature is a blend around 25-26\AA{} attributed to the O\textsuperscript{+6} He$\alpha$ lines at 21.6\AA{}, 21.8\AA{}, and 22.1\AA{} redshifted to the quasar frame.

The most prominent low ionization absorption lines are those of C, N, O and Ne. The neutral oxygen edge is clearly seen around 23-23.5\AA{}. The Ly$\alpha$ and He$\alpha$ (and some $\beta$ and $\gamma$) lines in the highly ionized absorber are also detected: C\textsuperscript{+4,+5}, N\textsuperscript{+5,+6}, O\textsuperscript{+6,+7} and Ne\textsuperscript{+7,+8}. Mg\textsuperscript{+9,+10} is also detected. Other ions are also detected, but with lower confidence.

Around 16-17\AA{} we marginally detect lines of Fe\textsuperscript{+8}-Fe\textsuperscript{+11}. These ions were also detected, very faintly, via emission by \cite{McCammon2002}. Lastly, in the range of 12-13\AA{} and 9-11\AA{} the model suggests highly ionized L-shell Fe and Ni respectively. These ions form at a temperature of $\sim10^7$ K, which diminishes our confidence in their detection, as there is no corroborating evidence for such high temperatures. Nevertheless, they cannot be ruled out on statistical grounds. The fact that no ions of Fe are seen between Fe\textsuperscript{+12} and Fe\textsuperscript{+18}, which should also exist in a 10\textsuperscript{7} K absorber, is  further indication that the presence of this absorber is questionable. We discuss the possibility of a higher temperature component in Section\,\ref{sec:PosHotAbsorber}.

\subsection{Column Densities} \label{subsec:ColDens}
Measured ionic column densities are presented in Table \ref{tab:colDens}. We find that the column densities of the dominant ions, O\textsuperscript{+6}, O\textsuperscript{+7} and Ne\textsuperscript{+8} are consistent with the column densities obtained by \cite{Fang&Xiaochuan2014} to within 90\%. The rest of the ions were not previously reported for 3C 273. The low ionization species, modeled with ISMAbs, reflect typical column densities of the galactic ISM \citep{Gatuzz2016}. Detections of upper limits are presented in Table \ref{tab:UpperLimitColDens}. In the remainder of the paper we focus on the high ionization species that are detected with high certainty and attributed to the local hot gas. 

\begin{table}
	\centering
	\caption{Ionic column densities of the ions detected using Ibifit and ISMAbs models} and their 90\% confidence intervals. \label{tab:colDens}
	\begin{threeparttable}	
		\begin{tabular}{lcl}
			\multicolumn{3}{c}{Ibifit} \\
			\hline
			\textbf{Ion} & \textbf{Column Density} & \textbf{Literature}\tnote{1} \\ 
			& (10\textsuperscript{16}cm\textsuperscript{-2}) & (10\textsuperscript{16}cm\textsuperscript{-2})\\ 
			\hline 
			C$^{+4}$ & $ 0.5_{-0.22}^{+0.24} $ \\ 
			C$^{+5}$ & $ 0.33_{-0.1}^{+0.11} $ \\ 
			N$^{+5}$ & $ 0.28_{-0.08}^{+0.09} $ \\ 
			N$^{+6}$ & $ 0.16_{-0.1}^{+0.12} $ \\ 
			O$^{+6}$ & $ 2.06_{-0.37}^{+0.42} $ &$ 2.3_{-0.5}^{+0.6}$\\ 
			O$^{+7}$ & $ 0.97_{-0.21}^{+0.24} $ &$ 0.59_{-0.22}^{+0.24}$\\ 
			Ne$^{+7}$ & $ 0.25_{-0.24}^{+0.29} $ &\\ 
			Ne$^{+8}$ & $ 0.86_{-0.32}^{+0.42} $ & $ 0.83_{-0.28}^{+0.75}$\\ 
			Ne$^{+9}$ & $ 0.8_{-0.48}^{+0.57} $ \\ 
			Mg$^{+9}$ & $ 1.12_{-0.6}^{+0.87} $ \\ 
			Mg$^{+10}$ & $ 0.77_{-0.55}^{+0.76} $ \\ 				
			Fe$^{+8}$ & $ 0.1_{-0.05}^{+0.05} $ \\ 
			Fe$^{+19}$ & $ 0.47_{-0.12}^{+0.12} $ \\ 
			Fe$^{+20}$ & $ 0.43_{-0.18}^{+0.2} $ \\ 
			Ni$^{+18}$ & $ 0.12_{-0.08}^{+0.09} $ \\ 					
			Ni$^{+20}$ & $ 0.07_{-0.02}^{+0.02} $ \\ 
			Ni$^{+21}$ & $ 0.05_{-0.02}^{+0.02} $ \\
			\hline
			\multicolumn{3}{c}{ISMAbs} \\			
			\hline
			C$^{+0} $ & $ 2.45_{-0.49}^{+0.3} $ \\ 
			N$^{+0} $ & $ 2.51_{-0.45}^{+0.26} $ \\ 
			N$^{+1} $ & $ 0.27_{-0.14}^{+0.11} $ \\ 
			N$^{+2} $ & $ 0.29_{-0.09}^{+0.12} $ \\ 
			O$^{+0} $ & $ 20.63_{-0.54}^{+0.37} $ \\ 
			O$^{+1} $ & $ 0.45_{-0.28}^{+0.31} $ \\ 
			O$^{+2} $ & $ 1.35_{-0.46}^{+0.24} $ \\ 
			Ne$^{+0} $ & $ 3.48_{-0.68}^{+0.36} $ \\ 
			Ne$^{+1} $ & $ <1.0 $ \\ 
			Ne$^{+2} $ & $ 0.43_{-0.27}^{+0.28} $ \\ 
			Mg$^{+0} $ & $ <0.44 $ \\
				
		\end{tabular}
		\begin{tablenotes}
			\item[1] \cite{Fang&Xiaochuan2014} - $1\sigma$ uncertainties
		\end{tablenotes}
	\end{threeparttable}
\end{table}

\begin{table}		
	\centering
	\caption{Upper limits of ionic column densities. Listed are the 90\% confidence intervals. \label{tab:UpperLimitColDens}}
		\begin{tabular}{lc}
			\multicolumn{2}{c}{Upper limits} \\
			\hline
			\textbf{Ion} & \textbf{Column Density} \\ 
			& (10\textsuperscript{16}cm\textsuperscript{-2}) \\ 
			\hline 
			O$^{+4} $ & $ <0.30 $ \\ 
			O$^{+5}$ & $ 0.14_{-0.13}^{0.14} $ \\				
			Mg$^{+8} $ & $ <1.51 $ \\ 
			Si$^{+7} $ & $ <31.0 $ \\ 
			Si$^{+8} $ & $ <43.1 $ \\ 
			Si$^{+9} $ & $ <68.2 $ \\ 
			Fe$^{+4} $ & $ <1.34 $ \\ 
			Fe$^{+5} $ & $ <0.1 $ \\ 
			Fe$^{+6} $ & $ <0.4 $ \\ 
			Fe$^{+7} $ & $ <0.12 $ \\
			Fe$^{+9} $ & $ <0.13 $ \\ 
			Fe$^{+10} $ & $ <0.14 $ \\ 
			Fe$^{+11}$ & $ 0.08_{-0.05}^{0.06} $ \\  
			Fe$^{+12} $ & $ <0.04 $ \\ 
			Fe$^{+13} $ & $ <0.02 $ \\
			Fe$^{+14} $ & $ <0.014 $ \\
			Fe$^{+15} $ & $ <0.032 $ \\
			Fe$^{+16} $ & $ <0.013 $ \\
			Fe$^{+17} $ & $ 0.0 $ \\
			Fe$^{+18}$ & $ 0.09_{-0.08}^{0.09} $ \\
			Ni$^{+19} $ & $ <0.05 $ \\			
		\end{tabular}	
\end{table}

\section{Absorption Measure Distribution (AMD)} \label{sec:AMD}
To characterize the elemental abundances, the hydrogenic column density and the temperature distribution of the hot absorber, we use the absorption measure distribution \citep[AMD, ][]{Holczer2007}. We define the $AMD$ for a coronal plasma as:
\begin{equation}
\label{AMDdef}
AMD\equiv\frac{\partial N_\mathrm{H}}{\partial \log(T)}
\end{equation}
where, $N_\mathrm{H}$ is the hydrogen column density and $T$ is the temperature. The measured ionic column densities can be expressed as an integral of the ionic abundances times the $AMD$ according to:
\begin{equation}
\label{AMDeq}
N_{Z^{+i}}=\int A_Z f_{Z^{+i}}(T) \frac{dN_\mathrm{H}}{d\log(T)} d\log(T)
\end{equation}
where $N_{Z^{+i}}$ is the column density in the element $Z$, ion $+i$, and $f_{Z^{+i}}$ is the ion $i$'s coronal fractional abundance \citep{Bryans2006}. $A_Z$ is the element $Z$ abundance relative to H.

\begin{figure}
	\includegraphics[width=\columnwidth]{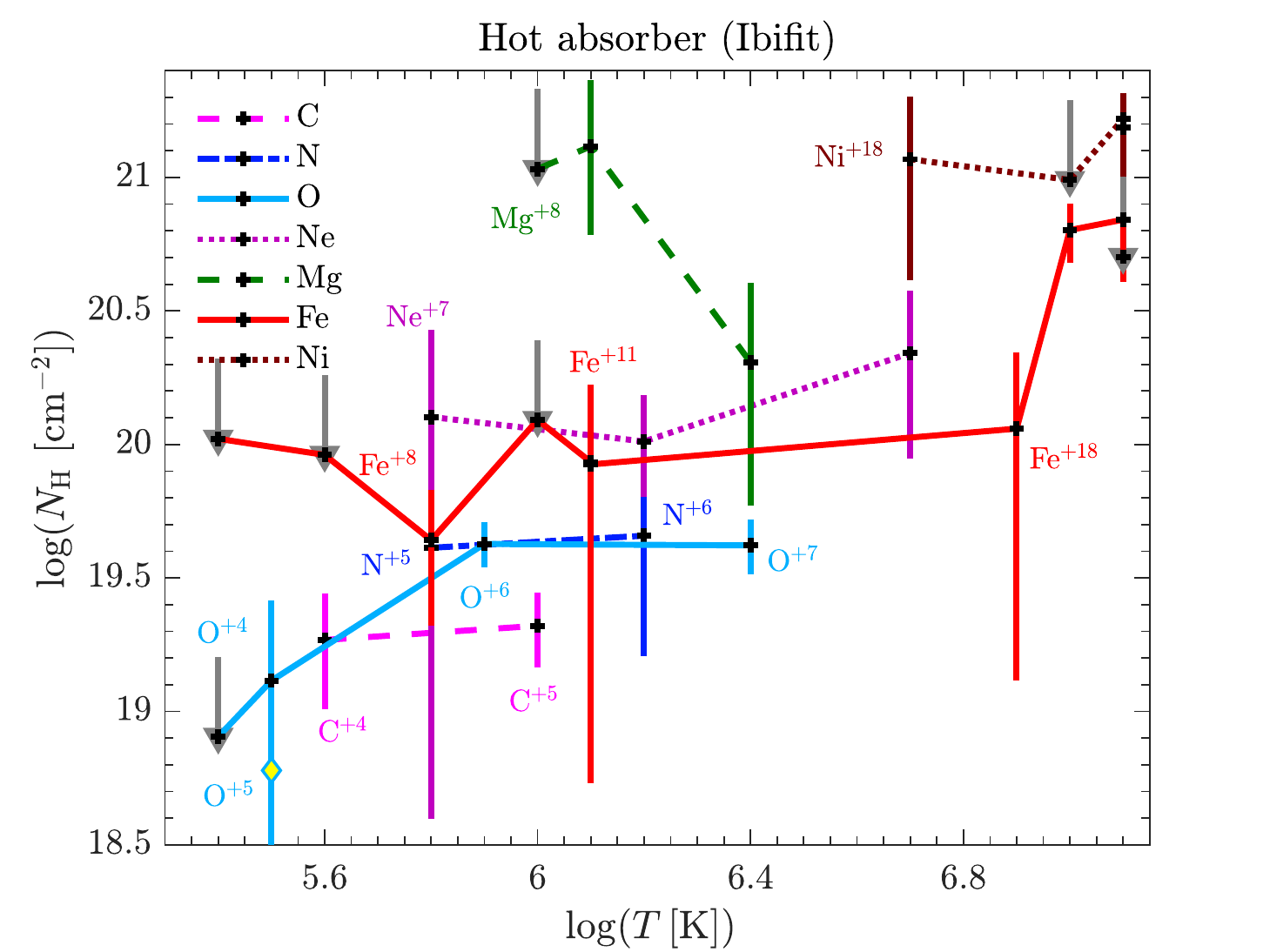}
	\caption{Equivalent $N_\mathrm{H}$ computed according to Eq. \ref{approxNH} as a function of $T_\mathrm{peak}$. Ion column densities and uncertainties are taken from Table \ref{tab:colDens}.  Vertical arrows represent upper limits (Table \ref{tab:colDens}). The diamond marker is the UV measurement \citep{Sembach2001}.}
	\label{fig:NHIons}
\end{figure}

As a first step, the $AMD$ can be approximated by assuming that each ion forms at the temperature of its peak fractional abundance, $f_{Z^{+i}}(T_\mathrm{peak})$. Assuming solar abundances \citep{Asplund2009}, $A_{Z_{\odot}}$, the equivalent hydrogen column density in each peak temperature from each ion is described as:
\begin{equation}
\label{approxNH}
N_\mathrm{H}=\frac{N_{Z^{+i}}}{f_{Z^{+i}}(T_\mathrm{peak}) A_{Z_{\odot}}}
\end{equation}
This is a lower limit for $N_\mathrm{H}$, since generally: $f_{Z^{+i}}(T)<f_{Z^{+i}}(T_\mathrm{peak})$. In Figure \ref{fig:NHIons} we plot the results of this estimate for each measured ionic column density (Table \ref{tab:colDens}). We also add the UV measurement of O\textsuperscript{+5} from \cite{Sembach2001}. 

The vertical shifts between elements in Figure \ref{fig:NHIons} could indicate non-solar abundance ratios. If we assume that oxygen is solar, other elements which are located above it in Figure \ref{fig:NHIons} must be over-abundant with respect to oxygen. To estimate the relative elemental abundances more precisely, we carry out the following procedure: We exploit the good measurements of O\textsuperscript{+6} and O\textsuperscript{+7} around 10\textsuperscript{6} K as a benchmark, and compute their error-weighted mean $N_\mathrm{H}$. C\textsuperscript{+4,+5}, N\textsuperscript{+5,+6}, Ne\textsuperscript{+7,+8}, Mg\textsuperscript{+10} and Fe\textsuperscript{+8,+10} all form around 10\textsuperscript{6} K. Their error weighted mean $N_\mathrm{H}$ relative to O in that regime yields their relative abundances $A_Z=\langle{N_\mathrm{H}^Z}\rangle/\langle{N_\mathrm{H}^O}\rangle$. The abundance uncertainties are Gaussian-propagated from the ionic column density uncertainties. 

For the higher temperatures, $6.7<\log(T)<7.3$, we compute the Ni weighted mean $N_\mathrm{H}$ from Ni\textsuperscript{+20,+21} and compare it to that of the high charge states of Fe\textsuperscript{+20,+21}. Assuming the Fe/O abundance ratio obtained above for 10\textsuperscript{6} K, we can calculate the Ni/O abundance as well. The resulting abundances are listed in Table \ref{tab:Abundances}, where they are compared to the abundances computed from the global fit to the spectrum with the Xspec Hotabs model described in Section\,\ref{sec:GlobalFitting}.

\begin{figure}
	\includegraphics[width=\columnwidth]{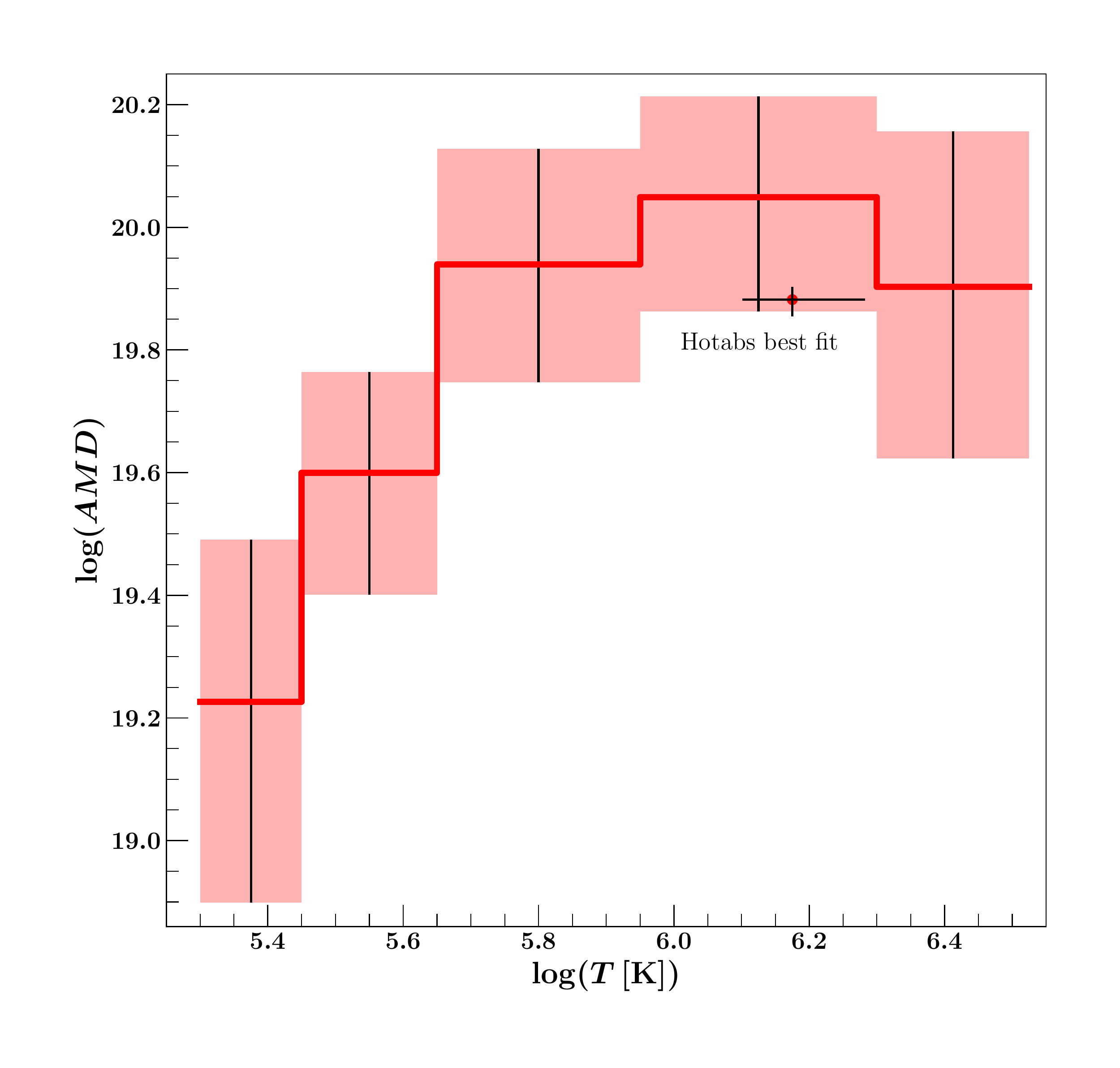}
	\caption{AMD result. The single data point is the Xspec one-temperature Hotabs fit and its 90\% confidence error bars. The overlap of the $\log(T)\simeq$6.2 bin with this point is a strong support for the existence of a $10^{6.2}$ K plasma.}
	\label{fig:AMD}
\end{figure}

Using the above results for $A_Z$, we can solve the set of $AMD$ equations (Eq. \ref{AMDeq}) without using the approximation of $f_{Z^{+i}}(T_\mathrm{peak})$. Each ion gives one such equation. We assume a discrete step-function $AMD$ of a few temperature bins, and solve simultaneously for all ions to obtain $AMD(T)$. The general process for solving the equations is described in \cite{Peretz&Behar2017}. The $AMD$ was calculated here using the ion detections from Table \ref{tab:colDens} and not upper limits, except for O\textsuperscript{+5} that we used from the UV detection \citep{Sembach2001}.  The $AMD$ results are plotted in Figure \ref{fig:AMD}. The 1T global-fitting result (Hotabs) described in the next section is superimposed on the $AMD$. The agreement between the two models strengthens our confidence in the process described in Section~\ref{sec:AMD} at least for the main temperature of 10\textsuperscript{6.2} K. The column density integrated over the AMD yields $N_\mathrm{H}\approx5\times10^{19}$ cm\textsuperscript{-2}.

\section{Global Modeling} \label{sec:GlobalFitting}
We compared our results with a global fitting model using Hotabs within the Xspec fitting suite \citep{Arnaud1996}. Hotabs does not fit each ion independently but is a physical collisional-ionization absorption model that fits the temperature and column density of the multi-element absorber by minimizing $\chi^2$ for the $\sim$2600 degrees of freedom (d.o.f). The fractional ionization abundances are calculated assuming a coronal plasma approximation. The global model used in Xspec is: (powerlaw+bbody+photemis)*ISMAbs*Hotabs, where Hotabs replaces Ibifit from before. We froze the oxygen abundance to its solar value, fitting all abundances relative to solar oxygen. The best-fit model is indistinguishable from the Ibifit one presented in Figure \ref{fig:spectrum}. 

\subsection{Abundances} \label{subsec:abundances}
The result of the elemental abundances from the global Hotabs fit compared to the individual-ion Ibifit are presented in Table \ref{tab:Abundances}. The uncertainties are given for the 90\% confidence level. 
\begin{table}
	\centering
	\caption{Relative abundances of the observed elements in the two models. \label{tab:Abundances}}
	\begin{tabular}{lccc}
		\hline
		\textbf{Ion} & \textbf{Hotabs} & \textbf{Ibifit} & \textbf{$\eta$} \\ 
		\hline
		C & $ 0.7_{-0.3}^{+0.3} $ & $ 0.5_{-0.3}^{+0.3} $ & 0.25 \\ 
		N & $ 1.0_{-0.3}^{+0.4} $ & $ 1.0_{-0.6}^{+0.6} $ & 0 \\ 
		Ne & $ 1.3_{-0.4}^{+0.6}  $ & $ 2.6_{-1.3}^{+1.3} $ & 0.9 \\ 
		Fe & $ 0.6_{-0.2}^{+0.2} $ & $ 1.4_{-0.7}^{+0.7} $ & 1 \\
		Ni & $ 0.0_{-0.0}^{+3.0} $ & $ 3.6_{-0.6}^{+0.6} $ & 1.2 \\ 
		\hline
		$ \chi^2/\mathrm{d.o.f} $ & $ 1.32 $ & $ 1.23 $ &  \\ 			
	\end{tabular}
\end{table}
The $\eta$ value in the table is the difference between the two values in terms of their statistical uncertainties, computed as: $\eta=|A_{Ibifit}-A_{Hotabs}|/\sqrt{\Delta A_{Ibifit}^2+\Delta A_{Hotabs}^2}$. The elemental abundances agree within 90\% confidence ($\eta\leqslant1$). Mg tends to be super-solar ($>3$) in both models due to the deep absorption lines (see Figure \ref{fig:NHIons}), which we find puzzling. The large difference in column density between Mg\textsuperscript{+9} and Mg\textsuperscript{+10} in Figure \ref{fig:NHIons} is also implausible since they are produced at approximately the same temperature. Therefore, we do not list Mg in the table, despite the significant spectral absorption at 9.16\AA{} and 9.29\AA{}. The agreement of the $AMD$ derived abundances (Section \ref{sec:AMD}) with Hotabs abundances suggests that the physical interpretation of a collisionally ionized absorber is valid.

In Table \ref{tab:comparisonWilliams} we compare the present measured abundances to those of \cite{Williams2005} who measured C, N and Ne in the Mrk 421 sight-line ($l=179.83^\circ, b=65.04^\circ$) assuming a galactic halo at a temperature of 10\textsuperscript{6.1}-10\textsuperscript{6.2} K, which are the same temperatures we measure for 3C 273. Our results are in agreement with \cite{Williams2005} despite the different sight-lines, however, our abundances are more tightly constrained, owing to the high S/N. If both absorbers originate in a galactic halo, the consistent temperature and abundances suggest a uniform halo \citep[e.g.,][]{Fang2015}.

\begin{table}
	\centering
	\caption{Comparison of the present elemental abundances with those of Mrk 421 \citep{Williams2005} \label{tab:comparisonWilliams}}
	\begin{tabular}{lcc}
		\hline
		\textbf{A\textsubscript{Z}/A\textsubscript{O}} & \textbf{3C 273 (Present work)} & \textbf{Mrk 421} \\ 
		\hline
		C/O  & $ 0.4-1.0 $ & $ 0.25-2 $  \\ 
		N/O  & $ 0.7-1.4 $ & $ <2.5 $  \\ 
		Ne/O & $ 0.9-1.9  $ & $ 0.25-4 $  \\ 
		Fe/O & $ 0.4-0.8 $ & $ --- $  \\
	\end{tabular}
\end{table}

\subsection{Broad emission lines}
The emission line component of our model indicates the presence of broad O and Ne lines. The global fitting model using Photemis in Xspec yields an ionization parameter of $\log\xi= 1.3\pm0.1$ ($\xi$ in erg\,s$^{-1}$\,cm), velocity broadening corresponding to $v_\mathrm{trub}=6400\pm1500$ km\,s$^{-1}$. This result is consistent with previous UV measurements which indicate gradually increasing width for higher charge states, e.g. $\sim6000$ km\,s$^{-1}$ (FWHM) for O\textsuperscript{+5} \citep{Laor1994}. The broad line component is highly significant. Adding it reduces $\chi ^2$ over the entire spectrum by 160 ($>$95\% confidence). 
The most prominent broad lines are the O\textsuperscript{+6} He$\alpha$ complex redshifted to 25.0-25.6\AA{}. Weaker O\textsuperscript{+7} and Ne\textsuperscript{+8} broad lines also appear as weak bumps at 22.0\AA{} and 15.9\AA{}, respectively (see Figure \ref{fig:spectrum}). The present X-ray lines are broader than Pa$\alpha$, which was recently measured at an unprecedented precision \citep[2000 km\,s$^{-1}$ at 150 light days,][]{Sturm2018}. This indicates they are closer in by a factor of $\sqrt{v_{X}/v_{Pa\alpha}}\approx1.8$), and are about 80 light days away from the black hole. 

\subsection{The possibility of a 10\textsuperscript{7} K absorber} \label{sec:PosHotAbsorber}
As mentioned in Section \ref{subsec:SpectralFitting}, the RGS spectrum cannot statistically rule-out the existence of highly ionized species of Ni and Fe in the 9-11\AA\ and 12-13\AA{} bands, respectively. As opposed to Ibifit, Hotabs takes into account mutual abundances of different elements, and therefore cannot fit some elements that are consistent with a 10\textsuperscript{7} K component while other elements are not. To examine this further, we added another Hotabs component to the Xspec model. i.e.: (powerlaw+bbody+photemis)*ISMAbs*Hotabs*Hotabs. We tied all the relative abundances, and fitted $N$\textsubscript{H} and $T$. The result was a negligible improvement in the reduced $\chi^2$ of 0.01. 
The most significant improvement of the two temperatures model is the appearance of the highly ionized Ni and Fe and the deepening of the O\textsuperscript{+7} Ly$\alpha$ line at 18.97\AA{} with some minor improvement of the N\textsuperscript{+5} He$\alpha$ at 28.78\AA{} and C\textsuperscript{+5} Ly$\alpha$ at 33.7\AA{} as well. At $T\sim10^7$ K the fractional abundance of O\textsuperscript{+7} is smaller than at $T\sim10^6$ K and at the tail of the ionic fraction curve. Consequently, this component requires high column density to account for the additional EW. On the other hand, at $T\sim10^7$K, Mg\textsuperscript{+11} is abundant and the Ly$\alpha$ line at 8.4\AA{} should be detected. The fact that we see no feature at 8.4\AA{} likely means that the 10\textsuperscript{7} K component does not exist. This conclusion is also supported by the lack of 10\textsuperscript{7} K emission lines in comprehensive sky surveys \citep{Henley&Shelton2010,Henley&Shelton2012}.

\section{Distance and Density} \label{sec:DistDensEst}
From combining the column densities with the emission measure of the same ions, one can extract the distance and density of the gas. Following \cite{Rasmussen2003}, we use the emission data from \cite{McCammon2002}; though McCammon's experiment was directed to the galactic coordinates: $l=90\degr, b=60\degr $ and 3C 273 is at: $l=289.95\degr, b=64.36\degr$.	Assuming a uniform spherical emitting volume, the flux measured from a bound $i \rightarrow j$ resonance transition is:
\begin{equation}
\label{Flux}
I_{ji}~[ \mathrm{ph\, cm^{-2}\, s^{-1}\, sr^{-1}}] \approx\frac{1}{4\pi}\int_{0}^{L}{n_i\, n_e\, Q_{ij}(T)\, dl}
\end{equation}
where $L$ is the radius of the sphere, Q\textsubscript{ij} is the collisional rate coefficient for electron impact excitation from lower level $i$ to higher level $j$, $n_i$ is the ionic density in the lower level, and $n_e$ is the electron density. We assume here a coronal plasma that is optically thin to its own emission, and treat the rate equations as a two-levels $(i,j)$ system in steady-state, i.e. the spontaneous emission rate ($A_{ji}$) exactly balances the collisional excitation rate:
\begin{equation}
\label{TwoLevelPopBalance}
\begin{array}{lcl}
n_jA_{ji} & = & n_in_eQ_{ij} \\
n_i+n_j & = & n_{Z^{+i}}
\end{array}
\end{equation}
where $n_{Z^{+i}}$ is the total number density of ion $Z^{+i}$. Then the ground level population can be estimated as:
\begin{equation}
\label{GroundStatePop}
n_i=\frac{n_{Z^{+i}}}{\Big(1+\frac{n_eQ_{ij}}{A_{ji}}\Big)}
\end{equation}

The ionic density can be re-written as a function of the relative abundances and the hydrogenic density:
\begin{equation}
\label{EqnziAsnH}
n_{Z^{+i}} \, n_e = \frac{n_{Z^{+i}}}{n_Z} \, \frac{n_Z}{n_\mathrm{H}} \, n_\mathrm{H} \, \frac{n_e}{n_\mathrm{H}} \, n_\mathrm{H}
\end{equation}
Where $n_{Z^{+i}}/{n_Z}(T)$ is the fractional ion abundance \citep{Bryans2006}, which is a function of temperature. $n_Z/n_\mathrm{H}=A_Z$ is the relative elemental abundance. $n_e/n_\mathrm{H}\approx1.2$ in a fully ionized solar-abundance plasma. Solving Eq. \ref{Flux} for a uniform density, we get:
\begin{equation}
\label{EqDist1}
L \, n_\mathrm{H}^2=\frac{4\pi \, I_{ji}}{\frac{n_{Z^{+i}}}{n_Z} \, \frac{n_Z}{n_\mathrm{H}} \, \frac{n_e}{n_\mathrm{H}} \, Q_{ij}}
\end{equation}
The ionic column density we measured is simply:
\begin{equation}
\label{EqColDens}
N_{Z^{+i}} = n_{Z^{+i}} \, L = n_\mathrm{H} \, \big( \frac{n_{Z^{+i}}}{n_z} \, \frac{n_z}{n_\mathrm{H}} \big) \, L
\end{equation}
Combining the two equations we derive the radius of the sphere:
\begin{equation}
\label{EqDistance}
L =\frac{\frac{n_e}{n_\mathrm{H}} \, Q_{ij}(T) \, N_{Z^{+i}}^2}{4 \pi \, I_{ji} \, \frac{n_{Z^{+i}}}{n_Z}(T) \, \frac{n_Z}{n_\mathrm{H}} }
\end{equation}

\begin{figure}
	\includegraphics[width=\columnwidth]{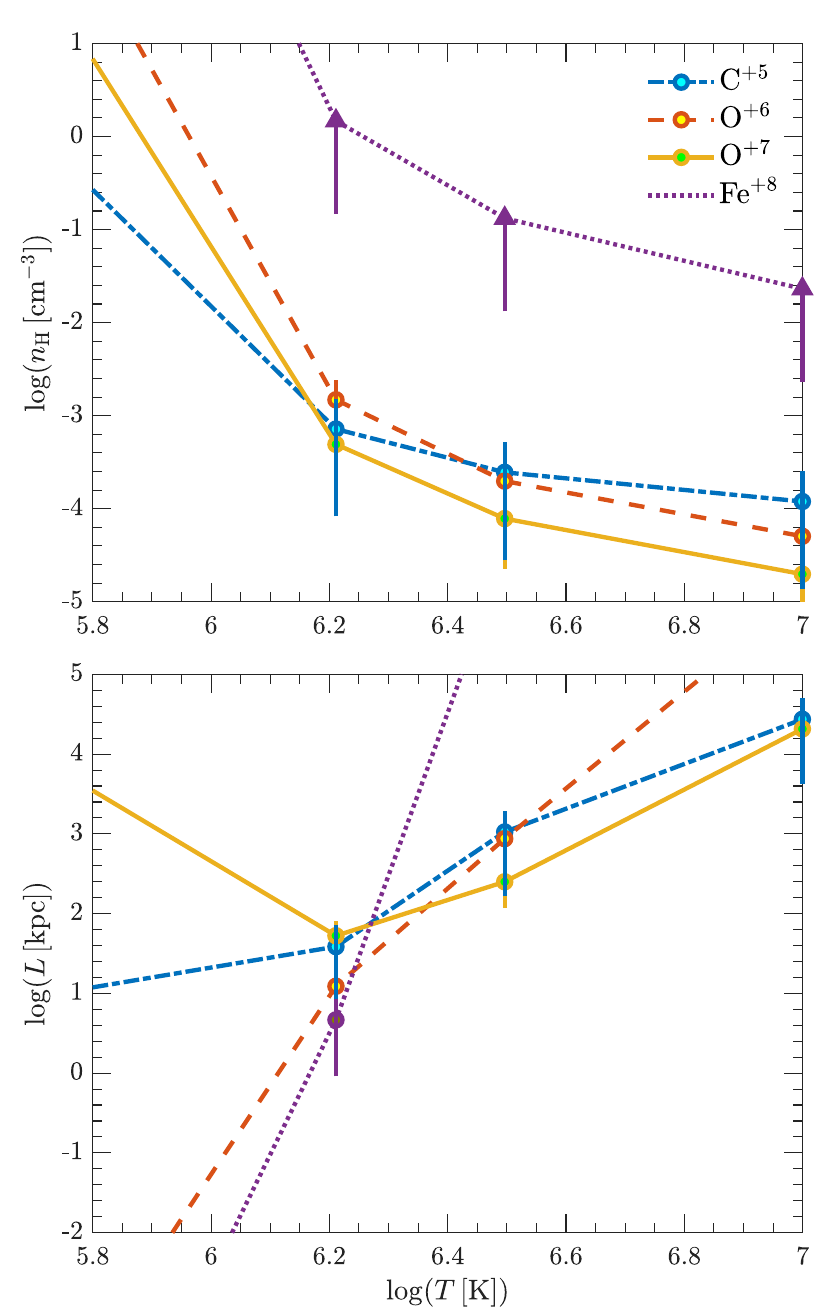}
	\caption{$L$ and $n$\textsubscript{H} computed as a function of temperature according to Eqs. \ref{EqDistance} and \ref{EqDist1}}
	\label{fig:DistAndnH}
\end{figure}

We compute $Q_{ij}(T)$ using the HULLAC atomic code \citep{Barshalom2001} for McCammon's four emission features. The results of $n$\textsubscript{H} and $L$ between $\log(T)=6-7$ are presented in Figure \ref{fig:DistAndnH}. These are listed explicitly in Table \ref{tab:Distances}. Given the temperature formation of these ions, we deem the distance and density at $\log(T)=6.2$ to be the most reliable.

\begin{table}
	\centering
	\caption{Distance $L$ (kpc) and densities $n$\textsubscript{H} (cm\textsuperscript{-3}) computed at $\log(T)=6.2$. \label{tab:Distances}}
	\begin{threeparttable}
		\begin{tabular}{lclll}	
			\hline		
			\textbf{Ion} & \boldmath{$\lambda$} & \textbf{Flux}\tnote{1} & \textbf{$L$} & \boldmath{$n$\textsubscript{H}} \\ 
			& (\AA{})& $\mathrm{(ph\, cm^{-2}\, s^{-1})}$& (kpc)& $(\mathrm{10^{-3}\, cm^{-3}})$ \\ 
			\hline
			C$^{+5}$ & $33.73$ & $ 5.4\pm 2.3$ & $38.6\pm32.5$ & $ 0.7\pm 0.8$ \\   
			O$^{+6}$ & $21.6$ & $ 4.8\pm 0.3$ & $12.3\pm 5.5$ & $ 1.5\pm 0.9$ \\ 
			O$^{+7}$ & $18.9$ & $ 1.6\pm 0.4$ & $52.8\pm28.1$ & $ 0.5\pm 0.3$ \\ 			
			Fe$^{+8}$ & $172-179$ & $100.0\pm50.0$ & $ 4.7\pm 3.7$ & $<1400$ \\ 
		\end{tabular}
		\begin{tablenotes}
			\item[1] \citep{McCammon2002}
		\end{tablenotes}
	\end{threeparttable}
\end{table}

From the C and O lines the absorbing (emitting) plasma is at a range of 1--80 kpc with a density of $2\times 10^{-4}-2\times10^{-3}$\,cm\textsuperscript{-3}, which places it in or just outside the hot galactic halo, see e.g. \citet{Sanders2004}. Accounting for the integrated AMD of $N_\mathrm{H}\approx5\times10^{19}$ cm\textsuperscript{-2} (see Section \ref{sec:AMD}, Figure \ref{fig:AMD}) these densities yield distance of 8--80 kpc. This distance range is in good agreement with the literature. \cite{Rasmussen2003} for $\log(T)=6.2$ obtain $L=40$ kpc and $n$\textsubscript{H}=5$\times10^{-4}$. \cite{Fang&Xiaochuan2014}, who used the \cite{Henley&Shelton2012} line fluxes, obtained a distance of 5-15 kpc. Using the \cite{McCammon2002} fluxes would increase this distance by a factor of $\sim3$. The present density is also in good agreement with \cite{Miller2016} who use a $\beta$ model for the density profile of the galactic halo. This density scales from $10^{-3}-10^{-4}$ between $L=10-100$ kpc. 

More generally, the galactic halo has a distribution of densities and temperatures with distance. If the density follows a $\beta$-profile, $n\propto r^{-3\beta}$, then the total mass $M(r)\propto r^{3-3\beta}$. At hydrostatic equilibrium, $T\propto M(r)/r\propto r^{2-3\beta}$ which in turn implies that $AMD\propto T^{(1-3\beta)/(2-3\beta)}$. Therefore, a proper measurement of the AMD could yield the value of $\beta$. In the present AMD measurement (Figure \ref{fig:AMD}) the slope of the $AMD$ with $T$ is only poorly constrained and does not provide a meaningful $\beta$ value.

The Fe feature yields lower distance and higher density than C and O though with high uncertainty. The emission is in fact a mix of Fe\textsuperscript{+8}-Fe\textsuperscript{+10} where we computed only the transition of Fe\textsuperscript{+8}. These M-shell ions form over a narrow temperature range and below $\log(T)=6.2$. Their sharp dependence on $T$, and their low measured abundance result in high estimated $L$ and $n$\textsubscript{H} values, with large uncertainties. Distances that are larger than 100 Mpc, as interpreted at $\log(T)\sim7$, would have been measured with redshift, but they are not shifted and are therefore ruled out.

\section{The line of sight towards 3C 273 in the UV} \label{sec:UVLineOfSight}
The line of sight towards 3C 273 contains supernova remnants (SNR) that may contaminate the spectrum and thus diminish the credibility of this sightline as a legitimate diagnostic of the galactic halo \citep{Williams2007, Nicastro2016}. Objects such as radio loops I and IV are thought to imply an old SNR that could emit X-rays, and the North Galactic Spur might also emit hot X-rays. To estimate the influence of the alleged SNR on the spectrum we examined the known UV systems detected in the 3C 273 sight-line \citep{Sembach2001, Sembach2003}.

The strongest absorption system at $z=0$ ($v$\,=\,-100\,--\,100 km/s) has a column density of $\log(N_{\mathrm{O}^{+5}} [\mathrm{cm}^{-2}])=14.77\pm0.04$ derived from FUSE or $14.85\pm0.1$ derived from OREFUS-II and is likely the system we detect in the X-ray. In the RGS spectrum we have an upper limit for O\textsuperscript{+5} of $\log(N_{\mathrm{O}^{+5}} [\mathrm{cm}^{-2}])\le15.4$, which is consistent with the UV. The O\textsuperscript{+6} column density however, is 40 times higher. 
The next strongest system is slightly redshifted, at $v$ = 100--240 km\,s$^{-1}$ and is attributed to radio loop IV \citep{Sembach2001}, but its column density is another order of magnitude lower, making it too weak to be detected by the RGS, thus unlikely to be the origin of the presently detected O\textsuperscript{+6} absorption. 

In addition to the $z=0$ absorber, there are eight different intergalactic systems with redshifts of 0.003 -- 0.14. All of them have low column densities of $\log(N_{\mathrm{O}^{+5}} [\mathrm{cm}^{-2}])<13.8$, which is $\sim$1 -- 2
orders of magnitude less than our measured column density of O\textsuperscript{+6}, yet again below our detection limit. To check whether we see absorption from any of these systems, we estimate the uncertainty of the redshift in our $z=0$ absorber. We constrain $z=2.2_{-2.2}^{+7.3} \times 10^{-5}$ in the Hotabs model. This value is well inside the galactic absorption detected in the UV and therefore we conclude that any contribution from a higher velocity system is negligible, either intergalactic or the radio loop IV component.

\section{Discussion and conclusions} \label{sec:DiscAndConc}
We analyzed the absorption X-ray spectrum towards the blazar 3C 273. Spectral lines were found at $z=0$ implying local absorbing gas. Cold and mildly ionized features, such as the neutral O K-edge were detected and attributed to the galactic disk cold neutral metal medium, as they are observed in other directions including towards galactic X-ray binaries \citep{Gatuzz2014}.

Absorption lines in highly ionized species were observed. This absorber is attributed to the galactic halo, surrounding the MW. The ionic column densities measured based on the most dominant absorption lines are in good agreement with previous literature where it exists \citep{Fang&Xiaochuan2014}. 

We measured the elemental abundances of C, N, Ne, and Fe relative to O in the local absorber towards 3C 273. The results are consistent with those of \cite{Williams2005} for Mrk 421 but better constrained. C/O, N/O and Ne/O are consistent with solar ratios, whereas the less constrained Fe is sub-solar.  We also reconstruct the AMD, that is the distribution of $N_\mathrm{H}(T)$, and find a dominant component at $\log(T)=6-6.2$. Absorption features such as those of L-shell ions of Fe and Ni from higher temperature of $\sim10^7$ K were marginally detected. The AMD method will be more useful for measuring the temperature distribution and even the density profile, when higher quality spectra will be available from future X-ray grating missions such as Arcus. The Hotabs global model yields a best fit of $\sim10^{6.2}$ K also in good agreement with the literature \citep{Fang&Xiaochuan2014, McCammon2002, Miller2016}. Adding a second Hotabs component at $10^7$ K does not improve the fit.

Using a wide-field soft X-ray emission line spectrum from \cite{McCammon2002} we estimated the distance and density of the emitting-absorbing local gas. We find that C\textsuperscript{+5}, O\textsuperscript{+6}, O\textsuperscript{+7} originate at distances of 1-80 kpc placing it in the, or just outside the MW halo, again, in good agreement with the literature \citep{Rasmussen2003, Fang&Xiaochuan2014}. 

Finally, emission features in the RGS spectrum revealed for the first time the broad line region of 3C 273 in X-rays, with velocities of 6400$\pm$1500 km\,s$^{-1}$.

\section*{Acknowledgements}
This work was supported by the Israel PAZY foundation grant \textit{Laboratory-astrophysics – Cold absorption}. We thank the referee for his useful comments.






\bsp	
\label{lastpage}
\end{document}